\documentclass[12pt]{article}

\usepackage{amsmath,amssymb,amsthm}

\newcommand{\EQUA}{\begin{equation}}
\newcommand{\EQQN}{\end{equation}}
\newcommand{\EQN}{\begin{eqnarray}}
\newcommand{\ENN}{\end{eqnarray}}

\newcommand{\I}{i}

\newcommand{\DE}{\Delta}
\newcommand{\G}{\gamma}

\newcommand{\HA}{{1 \over 2}}

\begin{document}

\title{\bf  Entropy Burst from Parabolic Potentials}

\author{ Tsunehiro Kobayashi\footnote{E-mail: 
kobayash@a.tsukuba-tech.ac.jp} 
and Toshiki Shimbori\footnote{E-mail: 
shimbori@het.ph.tsukuba.ac.jp} \\
{\footnotesize\it $^*$Department of General Education 
for the Hearing Impaired,}
{\footnotesize\it Tsukuba College of Technology} \\
{\footnotesize\it Ibaraki 305-0005, Japan}\\
{\footnotesize\it $^\dag$Institute of Physics, University of Tsukuba}\\
{\footnotesize\it Ibaraki 305-8571, Japan}}

\date{}

\maketitle

\begin{abstract}
The change of the energy of ground state is investigated in 
a thermodynamical process 
by using the model described by 
one-dimensional harmonic oscillator + two-dimensional isotropic parabolic 
potential barrier such as 
$V(x,y,z)=m\omega^2 x^2/2 -m\gamma^2 (y^2+z^2)/2$. 
In the process where two independent many-particle systems  suddenly 
touch with each other, 
it is shown that the lowest energy after the interaction can possibly be 
smaller than that before the interaction and then 
the entropy burst can occur. 
\end{abstract}

\thispagestyle{empty}

\setcounter{page}{0}

\pagebreak

In the previous paper~\cite{sk4} it has been shown that even-dimensional 
parabolic potential barriers such as 
$V(y,z)=-m\gamma^2 (y^2+z^2)/2$ 
have infinitely degenerate stationary-states 
with the zero real energy eigenvalue, 
which are described by stationary flows round the center of the potential. 
This result is really surprising, because in general 
the parabolic potential barriers have been 
considered as the model for unstable states~[2-7].  
In fact it is shown that one has only unstable states like resonances 
in one-dimensional parabolic potential barriers like 
$V(x)=-m\gamma^2 x^2/2$, 
which are represented by the solutions of the conjugate space of 
Gel'fand triplet~\cite{bohm} with imaginary energy eigenvalues 
$\mp \I(n+1/2)\hbar\gamma\ (n=0,1,2,\cdots)$~[2-7]. 
The existence of such stationary states 
in even-dimensional is not only surprising fact 
but also suggests the existence of the stable many-body systems 
constructed from the stationary states. 
Following this line of consideration, we have shown that there exist stable 
many-body systems composed of the stationary flows in statistical mechanics 
including all states derived in the Gel'fand triplet formalism~\cite{ks5}. 
An important difference appears in the fact that 
in the new statistical mechanics a new freedom arises from that of 
the imaginary part of energy eigenvalues which is only allowed in the extended 
spaces of the Gel'fand triplet. 
Actually the thermodynamical probability $W$ 
is expressed by the product of 
the thermodynamical probability $W^\Re$ 
with respect to the real part of the energy 
and that $W^\Im$ 
with respect to the imaginary part of the energy 
such that 
$W=W^\Re W^\Im$. 
Therefore 
the entropy which is defined by 
$S=k_{\rm B}\log W$ becomes the sum of the usual Boltzmann 
entropy $S^\Re=k_{\rm B}\log W^\Re$ 
induced from the freedom of the real part of 
the energy eigenvalues and a new 
one $S^\Im=k_{\rm B}\log W^\Im$ originating from that of the imaginary part 
as $S=S^\Re+S^\Im$.
(In details, see ref.~9.) 
It is natural to expect that there exists some kind of entropy transfer 
between the two entropies $S^\Re$ and $S^\Im$. 

In this paper we shall study the entropy transfer in an explicit example 
by using a simple model 
which is described by 
one-dimensional harmonic oscillator (HO) 
+ two-dimensional isotropic parabolic potential barrier (PPB) such as 
\begin{equation}
V(x,y,z)=\HA m\omega^2 x^2 -\HA m\gamma^2 (y^2+z^2),
\end{equation} 
where $m$ is the mass of particle.
Note that the HO part is necessary for confirming the entropy transfer, 
because the PPB cannot have any freedom of the real 
energy eigenvalues, that is, the real energy eigenvalue is fixed at 0. 
The energy eigenvalues of the potential (1) are given by 
\begin{equation}
E_{n_x n_y n_z}=
\left(n_x+\HA\right)\hbar\omega \mp\I (n_y-n_z)\hbar\G
\end{equation}
for the solutions involving the stationary states, where 
the first and the second terms, respectively, denote the contributions 
from the HO and the PPB and 
$n_x,\ n_y,\ n_z=0,1,2,\cdots$. 
The stationary states appear for $n_y=n_z$, of which relation washes out 
the imaginary part from the eigenvalues~\cite{sk4}. 
It is now transparent that the infinite degeneracy of the all states having 
the energy eigenvalues of (2) originates from the freedom of the imaginary 
eigenvalues. 
Note that all stationary states are described by stationary flows round 
the center of the potential~\cite{sk4}. 
Actually the difference between the infinitely degenerate states 
is represented by the difference of the 
stationary flows.
Note that complex velocity potentials $W$ 
which are well-known in hydrodynamics 
can be introduced and $W=\pm \gamma (y+\I z)^2/2$ is obtained for 
the first few stationary eigenstates in 
the two-dimensional PPB $V(y,z)=-m\gamma^2 (y^2+z^2)/2$~\cite{sk4,sk3}. 

Let us consider a system being in a thermal equilibrium, 
which is composed of $N_1$ particles trapped by 
the potential 
\begin{equation}
V_1=\sum_{i=1}^{N_1}\left[\HA m\omega^2 x_i^2 
-\HA m\gamma^2 (y_i^2+z_i^2)\right].
\end{equation}
The  energy of the stationary ground state is uniquely determined 
by 
\begin{equation}
E_1=\HA N_1\hbar\omega. 
\end{equation}
This state, of course, has no degeneracy arising 
from the HO 
and then the Boltzmann entropy $S^\Re=0$. 
It, however, has a diverging entropy with respect to the new entropy $S^\Im$, 
since the infinite degeneracy arises from the PPB. 
It is noted here that the stable many-body systems in the PPB 
are described as the nets of stationary circular-flows, the two 
joints of which are not connected by the wavefunctions of 
the stationary eigenstates but they are 
connected by the stationary flows of the PPB~\cite{ks5}. 
This fact means that one can directly look in the net-structure (texture) 
composed of the stationary flows, because the flows 
are basically observable in quantum mechanics. 
It is, therefore, understood that the systems have some kind 
of classical property 
that the components inside of the systems are observable. 
Thus one sees that the appearance of the entropy $S^\Im$ originates from 
this semiclassical property of the stable many-body systems 
in the even-dimensional PPB. 

Now let us study a thermal process which is described by the mixing of 
two independent many-particle systems written by the potential given in (3). 
We consider the following situation; the two independent systems are 
suddenly put on their interactive region, where 
one has the center at $x=a$ and $y=z=b$ and the other at $x=-a$ and $y=z=-b$,
and then the mixing starts and they finally make a new stable system 
in a thermal equilibrium. 
Before the mixing 
the ground-state energy is uniquely determined by 
\begin{equation}
E_{\rm bef}=\HA (N_1+N_2) \hbar \omega,
\end{equation}
where $N_1$ and $N_2$ are, respectively, 
the number of the constituent particles of the system one 
and that of the other. 
After the mixing the two potentials being 
at the two centers  have the effects on 
the all constituent particles and then the potential is written by 
\begin{align}
V_{\rm T}=\sum_{i=1}^N \biggl\{
&\HA m\omega^2 \left[(x_i-a)^2+(x_i+a)^2\right] \notag \\
&-\HA m\gamma^2 \left[(y_i-b)^2+(z_i-b)^2+(y_i+b)^2+(z_i+b)^2\right]
\biggr\},
\end{align}
where $N=N_1+N_2$. 
This potential is rewritten as 
\begin{equation}
V_{\rm T}=\sum_{i=1}^N \left[ m\omega^2 x_i^2 - m\gamma^2 (y_i^2+z_i^2)\right]
     +Nm(\omega^2 a^2-2\G^2b^2).
\end{equation}
The differences between the potentials before  and after the mixing 
appear in the following two points: 
One is the fact that the potential curvatures 
after the mixing become twice as large as 
those before the mixing and the other is the appearance of 
a real constant term 
$Nm(\omega^2 a^2-2\G^2b^2)$. 
These effects can be seen in the ground-state energy such that 
\begin{equation}
E_{\rm aft}={1 \over \sqrt{2}} N \hbar \omega+Nm(\omega^2 a^2-2\G^2b^2).
\end{equation}
The difference between the ground-state energies before and after 
the mixing is evaluated as 
\begin{equation}
\DE E_0=\left({1 \over \sqrt{2}}-\HA\right) N \hbar \omega
+Nm(\omega^2 a^2-2\G^2b^2),
\end{equation}
where 
$\DE E_0=E_{\rm aft}-E_{\rm bef}$.
The difference can be negative, even if 
the first term of the difference is definitely positive. 
Namely, when the relation 
\begin{equation}
\left({1 \over \sqrt{2}}-\HA\right) \hbar \omega+m\omega^2 a^2
<2m\G^2 b^2
\end{equation}
is satisfied, the difference become negative. 
It should be noted that the negative contribution originates only 
from the two-dimensional PPB. 
One easily see that the difference always becomes positve, if 
the potential is described only by HO. 
The existence of the PPB is essential to derive the 
negative difference. 
The change of the potentials producing real energy eigenvalues 
before and after the mixing are illustrated in fig.~1. 
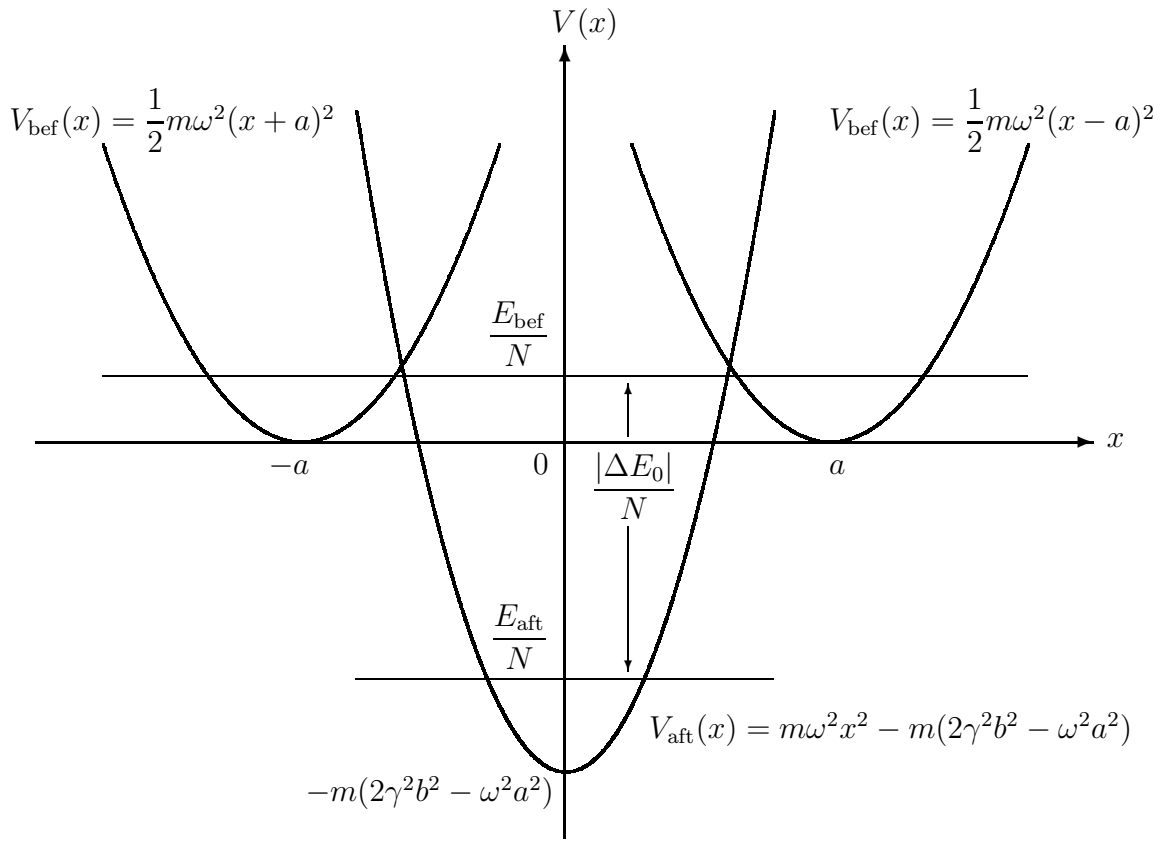
\begin{figure}[htbp]
\thicklines
\begin{center}
  \begin{picture}(400,300)
   \put(0,150){\vector(1,0){400}}
   \put(200,0){\vector(0,1){300}}
   \put(188,138){$0$}
   \put(405,148){$x$}
   \put(195,305){$V(x)$}
   
   \qbezier[800](225,262.5)(300,37.5)(375,262.5)
   \put(300,138){$a$}
   \put(300,268){$V_{\rm bef}(x)=\dfrac{1}{2}m\omega^2(x-a)^2$}

   \qbezier[800](25,262.5)(100,37.5)(175,262.5)
   \put(88,138){$-a$}
   \put(-10,268){$V_{\rm bef}(x)=\dfrac{1}{2}m\omega^2(x+a)^2$}
   
   \qbezier[8000](120.94,275)(200,-225)(279.06,275)
   \put(102,14){$-m(2\gamma^2b^2-\omega^2a^2)$}
   \put(232,38)
   {$V_{\rm aft}(x)=m\omega^2x^2-m(2\gamma^2b^2-\omega^2a^2)$}

   \thinlines
   \put(25,175){\line(1,0){350}}
   \put(170.72,187){$\dfrac{ E_{\rm bef}}{N}$}

   \put(120.94,60.36){\line(1,0){158.12}}
   \put(172,72.36){$\dfrac{ E_{\rm aft}}{N}$}

   \put(224,152){\vector(0,1){20}}
   \put(224,118){\vector(0,-1){54.64}}
   \put(210,129){$\dfrac{|\DE E_0|}{N}$}
  \end{picture}
\end{center}
\caption[]{Potentials producing real energy eigenvalues 
for one constituent. }
\label{fig:6.1}
\end{figure}
This real and negative energy $\DE E_0$ must be absorbed 
in the energy of the HO, 
because the PPB can only have 
the freedom of pure imaginary eigenvalues. 
This means that the ground-state energy before the mixing becomes 
an excited-state energy 
after the mixing. 
The excitation energy is evaluated by 
$
E_{\rm exc}=|\DE E_0|.  
$
Note here that $E_{\rm exc}$ is generally a macroscopic order because it is 
proportional to the total particle number $N$. 
Thus the system have the freedom arising from the real energy, which 
is given by the Boltzmann entropy 
\begin{equation}
S^\Re=k_{\rm B}\log W^\Re(E_{\rm exc}),
\end{equation}
where
$$
W^\Re(E_{\rm exc})={(M+N-1)! \over M!(N-1)!}
$$
and $M$ stands for the 
excitation number of the HO after the mixing, 
which is defined by the maximum integer being smaller than 
$E_{\rm exc}/\sqrt{2}\hbar\omega$. 
It is well-known that one can express the entropy as
\begin{equation}
S^\Re=k_{\rm B}\left[ (M+N)\log (M+N)-M\log M-N \log N \right],
\end{equation}
where $M,N\gg 1$ 
are postulated. 
The temperature is now given by 
\begin{equation}
T={\sqrt{2}\hbar\omega \over k_{\rm B}}
\left\{\log\left[{m(2\G^2b^2-\omega^2 a^2)
+(\sqrt{2}+1)\hbar\omega/2 \over 
m(2\G^2b^2-\omega^2 a^2)-(\sqrt{2}-1)\hbar\omega/2 }\right]
\right\}^{-1}. 
\end{equation}

We have shown that the entropy production 
from the PPB is possible. 
Namely, the mixing of two systems changes the ground state and then 
the energy of the ground state after the mixing can 
be much smaller than that before the mixing. 
The essential point of this process is the existence of stable states in the 
repulsive potentials like the PPB~\cite{sk4}. 
PPB's can possibly be good approximations to repulsive 
forces which are very week at the center of the force as same as 
the fact that HO's are known as good approximations 
to attractive forces being very week at the center. 
We can, therefore, expect that entropy productions will occur in 
real thermodynamical processes. 
One may understand that the entropy production is caused by 
the entropy transfer between $S^\Re$ and $S^\Im$, 
though one cannot directly show it in the present model 
because the entropies 
$S^\Im$'s 
in the initial and the final states are infinity. 

It will happen in realistic processes that the number 
$E_{\rm exc}/\sqrt{2}\hbar\omega(\geq M)$ is not integer. 
In such processes some part of energy must be emitted from the systems 
to make a stable system after the mixing. 
Sometimes the most of the energy $E_{\rm exc}$ will possibly be emitted and 
the system goes to the fatal ground state, 
since the change of the curvature of the PPB 
makes all stationary states in the initial state unstable in the final 
state. 
In such cases the observers 
will see 
the burst of energy, which will be seen as the burst of entropy also. 
This fact indicates that one can make macroscopic energy bursts, 
provided that one can prepare the thermally stable systems composed of 
stationary flows in PPB's. 
That is to say, there is a possibility of producing energies 
in purely thermal processes 
without any nuclear fusions  at ordinary temperatures.
Finally we would also like to note that it will be a very interesting trial 
to describe the birth of the Universe 
in the present scheme. 

\pagebreak

%


\begin{thebibliography}{99}
  \bibitem{sk4}
  T.~Shimbori and T.~Kobayashi, 
  {\it Stationary Flows of the Parabolic Potential Barrier 
  in Two Dimensions}, 
  quant-ph/0006019.
  
  \bibitem{barton}
  G.~Barton, 
  Ann. Phys. {\bf 166} (1986) 322. 
  
  \bibitem{bcd}
  P.~Briet, J.~M.~Combes and P.~Duclos,
  Comm. Partial Differential Equations {\bf 12} (1987) 201.
  
  \bibitem{bv}
  N.~L.~Balazs and A.~Voros, 
  Ann. Phys. {\bf 199} (1990) 123.
  
  \bibitem{cdlp}
  M.~Castagnino, R.~Diener, L.~Lara and G.~Puccini, 
  Int. J. Theor. Phys. {\bf 36} (1997) 2349, 
  quant-ph/0006011.
  
  \bibitem{sk}
  T.~Shimbori and T.~Kobayashi, 
  Nuovo Cim. {\bf 115B} (2000) 325. 
  
  \bibitem{s2}
  T.~Shimbori, 
  \emph{Operator Methods of the Parabolic Potential Barrier}, 
  quant-ph/9912073, to appear in Phys. Lett. {\bf A}. 
  
  \bibitem{bohm}
  A.~Bohm and M.~Gedella, 
  {\it Dirac Kets, Gamow Vectors and Gel'fand Triplets},
  Lecture Notes in Physics {\bf 348} (Springer, Berlin) 1989.
  
  \bibitem{ks5}
  T.~Kobayashi and T.~Shimbori, 
  {\it Statistical Mechanics for States with Complex Eigenvalues 
  and Quasi-stable Semiclassical Systems}, 
  cond-mat/0005237. 
  
  \bibitem{sk3}
  T.~Shimbori and T.~Kobayashi, 
  {\it ``Velocities'' in Quantum Mechanics}, 
  quant-ph/0004086. 
\end{thebibliography}
\end{document}